\documentclass[aps, prb, reprint, superscriptaddress, floatfix]{revtex4-1}
\usepackage{amssymb}
\usepackage{amsmath}
\usepackage{graphicx}
\usepackage{bm}

\begin{document}

\title{Plasmon modes of coupled quantum Hall edge channels in the presence
of disorder-induced tunneling}
\author{Toshimasa Fujisawa}
\email{fujisawa@phys.titech.ac.jp}
\affiliation{Department of Physics, Tokyo Institute of Technology, 2-12-1 Ookayama,
Meguro, Tokyo, 152-8551, Japan.}
\author{Chaojing Lin}
\affiliation{Department of Physics, Tokyo Institute of Technology, 2-12-1 Ookayama,
Meguro, Tokyo, 152-8551, Japan.}
\affiliation{Tokyo Tech Academy for Super Smart Society, Tokyo Institute of Technology,
2-12-1 Ookayama, Meguro, Tokyo, 152-8551, Japan.}

\begin{abstract}
Coupled quantum Hall edge channels show intriguing non-trivial modes, for
example, charge and neutral modes at Landau level filling factors 2 and 2/3.
We propose an appropriate and effective model with Coulomb interaction and
disorder-induced tunneling characterized by coupling capacitances and
tunneling conductances, respectively. This model explains how the transport
eigenmodes, within the interaction- and disorder-dominated regimes, change
with the coupling capacitance, tunneling conductance, and measurement
frequency. We propose frequency- and time-domain transport experiments, from
which eigenmodes can be determined using this model.
\end{abstract}

\date{\today }
\maketitle

\section{Introduction}

Integer and fractional quantum Hall systems provide unique opportunities for
studying many-body effects in two-dimensional topological insulators, where
chiral one-dimensional edge channels dominate the transport characteristics 
\cite{BookEzawa,BookWen}. Particularly, when multiple edge channels are
mutually coupled, the system can exhibit fractionalization into non-trivial
excitations \cite{BookWen,BookGiamarchi}. The phenomena can be understood by
considering the transport eigenmodes of the channels. For example,
copropagating integer channels along the edge of the quantum Hall state at
the Landau level filling factor $\nu =2$ can be understood with the spin and
charge (dipolar) modes, where Tomonaga-Luttinger physics show up \cite%
{Bocquillon-NatCom2013,Inoue-PRL2014,Hashisaka-NatPhys2017}. For another
important example of counterpropagating integer and fractional channels at $%
\nu =2/3$, similar charge and neutral modes can be studied with the edge
reconstruction in the hole-conjugate fractional state \cite%
{Bid-PRL2009,Sabo-NatPhys17,Lafont-Science2019}.

These examples have been studied by utilizing mesoscopic functional devices
such as quantum point contacts and quantum dots to reveal non-equilibrium
states and their dynamics \cite%
{Ji-Nature03,Kamata-PRB10,Altimiras-NatPhys10,Kamata-NatNano2014,Itoh-PRL18}%
. However, the above two cases have been studied separately as the
underlying mechanisms are largely different. While the spin and charge modes
at $\nu =2$ are formed by the Coulomb interaction between the channels \cite%
{Berg-PRL2009}, the charge and neutral modes at $\nu =2/3$ arise from
disorder-induced tunneling between the channels \cite%
{KaneFisher-PRB1995,KaneFisherPolchinski-PRL1994}. As a result, the
characteristics of the $\nu $ = 2 and $\nu $ = 2/3 modes are also different.
The spin and charge modes at $\nu =2$ cannot be detected at equilibrium with
conventional conductance measurements, but can be identified by using
methods such as high-frequency transport, shot noise detection,
energy-resolved measurements \cite%
{Bocquillon-NatCom2013,Inoue-PRL2014,Hashisaka-NatPhys2017}. On the other
hand, the charge and neutral modes at $\nu =2/3$ can be discussed with the
linear conductance that changes with the channel length \cite%
{Grivnin-PRL2014,CJLin-PRB2019}. In addition to these individually
successful investigations of each system, it would be advantageous to
coherently understand the two cases, because actual devices may involve both
interaction and tunneling. Indeed, a recent high-frequency experiment
supports the emergence of pure neutral and charge modes in the $\nu =2$
system in the presence of tunneling \cite{CJLin}. There may also be some
cases where disorder-induced scattering is suppressed in the $\nu =2/3$
case. Therefore, it is beneficial to consider both interaction and tunneling
in the theoretical model, as well as in the experimental analysis.

In this paper, we provide a plasmon scattering model for two parallel edge
channels in the presence of arbitrary interaction and tunneling occurring
between them, particularly focusing on the eigenmodes of the $\nu =2$ and $%
\nu =2/3$ cases. Using this model, the manner in which transport eigenmodes
change from the uncoupled regime to the interaction-dominated regime and the
disorder-dominated regime can be understood systematically by tuning the
parameters. The transport eigenmodes can be investigated experimentally by
measuring the scattering matrix elements of plasmons for a finite length of
the coupled edge channels. The model can be extended to multiple (more than
two) channels, as well as other 2D topological insulators where the coupling
of multiple channels plays an essential role.

\section{Plasmon model}

\begin{figure}[tbp]
\begin{center}
\includegraphics[width = 2.4in]{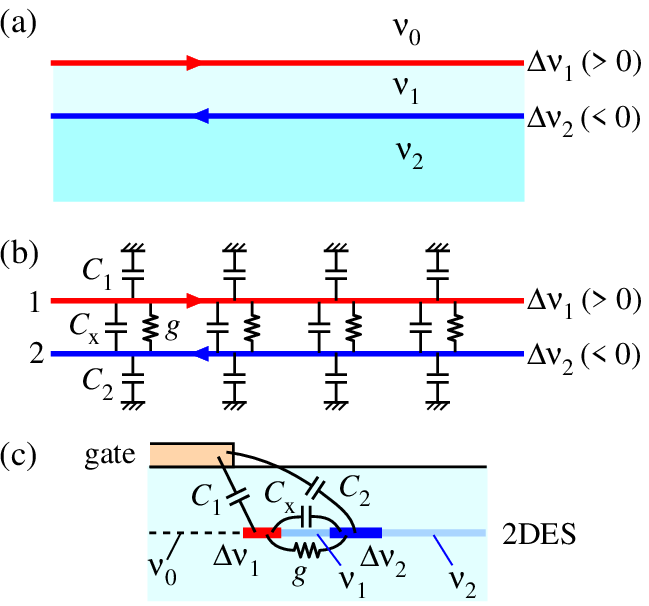}
\end{center}
\caption{(a) Schematic of the coupled edge channels with $\Delta \protect\nu %
_{1}=\protect\nu _{1}-\protect\nu _{0}$ and $\Delta \protect\nu _{2}=\protect%
\nu _{2}-\protect\nu _{1}$ formed along the boundaries of quantum Hall
states at $\protect\nu _{0}$, $\protect\nu _{1}$, and $\protect\nu _{2}$.
Righ-moving channel with $\Delta \protect\nu _{1}>0$ and left-moving channel
with $\Delta \protect\nu _{1}<0$ are illustrated. (b) Plasmon model with
distributed capacitances, $C_{1}$, $C_{2}$, and $C_{\mathrm{x}}$, and
distributed tunneling conductance $g$. (c) Realistic geometry of the coupled
edge channels formed along the perimeter of the gate. Two channels with $%
\Delta \protect\nu _{1}$ and $\Delta \protect\nu _{2}$ are formed along
boundaries of insulation regions with $\protect\nu _{0}$, $\protect\nu _{1}$%
, and $\protect\nu _{2}$.}
\end{figure}

We apply a plasmon scattering approach developed with one-dimensional
systems \cite%
{BookGiamarchi,Safi-PRB1995,SafiEurPhys,Hashisaka-PRB2012,Protopopov-AnnPhys2017}%
.We consider two chiral edge channels, $c=\left\{ 1,2\right\} $, along the
boundaries of the quantum Hall (insulating) regions with Landau level
filling factors, $\nu _{0}$, $\nu _{1}$, and $\nu _{2}$, as shown in Fig.
1(a). The Hall conductance $\sigma _{c}=\sigma _{q}\Delta \nu _{c}$ of
channel $c$ is given by the difference of the filling factor $\Delta \nu
_{c}=\nu _{c}-\nu _{c-1}$ in the neighboring regions and quantized
conductance $\sigma _{q}=e^{2}/h$. The sign of $\Delta \nu _{c}$ determines
the chirality; positive for right movers and negative for left movers.

For example, two copropagating integer channels, $\Delta \nu _{1}=\Delta \nu
_{2}=1$, which we shall call the $\nu =2$ case, are formed with $\nu _{0}=0$%
, $\nu _{1}=1$, and $\nu _{2}=2$ \cite{BookEzawa,Berg-PRL2009}. Two
counterpropagating integer/fractional channels, $\Delta \nu _{1}=1$ and $%
\Delta \nu _{2}=-1/3$, which we shall call the $\nu =2/3$ case, can be
studied with $\nu _{0}=0$, $\nu _{1}=1$, and $\nu _{2}=2/3$ \cite%
{MacDonald-PRL1990,Wen-PRL1990}. The interactions between the channels, as
well as those within each channel, can be described with mutual capacitance $%
C_{\mathrm{x}}$ and self capacitances $C_{\mathrm{1}}$ and $C_{\mathrm{2}}$
defined for a unit length, as shown in Fig. 1(b) \cite{Hashisaka-NatPhys2017}%
. The self capacitances are practically determined by capacitances to the
ground, such as nearby gate electrodes or a potential reference. The
disorder-induced tunneling can be described with tunneling conductance $g$
defined for the unit length \cite{Nosiglia-PRB2018}.

Such edge channels can be formed in a gated AlGaAs/GaAs heterostructure, as
shown in Fig. 1(c). The bulk of the 2DEG on the right side is prepared in a
QH state at $\nu _{2}$, and the application of a negative voltage on the
surface metal gate depletes the underlying 2DEG on the left side to set $\nu
_{0}=0$. A narrow QH state at $\nu _{1}$ is formed for specific choices of $%
\nu _{2}$ and $\nu _{0}$. Interactions for the system comprising the
channels, $\Delta \nu _{1}$ and $\Delta \nu _{2}$, are described with
geometric capacitances $C_{\mathrm{1}}$, $C_{\mathrm{2}}$, and $C_{\mathrm{x}%
}$ as illustrated. Tunneling between the channels is allowed in the presence
of impurity potential. Besides, spin-flip mechanisms, such as spin-orbit and
hyperfine interactions, may be required for the $\nu =2$ case. We focus on
the eigenmode in the presence of both interactions and tunneling.

The charge density $\rho _{c}$ per unit length and electrostatic potential $%
V_{c}$ of channel $c$ are related by%
\begin{equation}
\left( 
\begin{array}{c}
\rho _{1} \\ 
\rho _{2}%
\end{array}%
\right) =\left( 
\begin{array}{cc}
C_{\mathrm{1}}+C_{\mathrm{x}} & -C_{\mathrm{x}} \\ 
-C_{\mathrm{x}} & C_{\mathrm{2}}+C_{\mathrm{x}}%
\end{array}%
\right) \left( 
\begin{array}{c}
V_{1} \\ 
V_{2}%
\end{array}%
\right) ,  \label{EqrCV}
\end{equation}%
which can be written as $\mathbf{\rho }=\mathbf{CV}$ with corresponding
vector notations, $\mathbf{\rho }=\left( 
\begin{array}{c}
\rho _{1} \\ 
\rho _{2}%
\end{array}%
\right) $ and $\mathbf{V}=\left( 
\begin{array}{c}
V_{1} \\ 
V_{2}%
\end{array}%
\right) $, and the capacitance matrix $\mathbf{C}$. Here, $\mathbf{\rho }$
and $\mathbf{V}$ change with time $t$ and coordinate $x$. The
electrochemical potential $\mu _{c}=eV_{c}+\rho _{c}/D_{c}$ is further
raised by the second term with the density of states $D_{c}$ \cite%
{Gabelli-PRL2007,Hashisaka-RevPhys2018}. By using its matrix form $\mathbf{D}%
=\left( 
\begin{array}{cc}
D_{1} & 0 \\ 
0 & D_{2}%
\end{array}%
\right) $, the vector form of $\mathbf{\mu }=\left( 
\begin{array}{c}
\mu _{1} \\ 
\mu _{2}%
\end{array}%
\right) $ is given by%
\begin{equation}
\mathbf{\mu }=\left( e\mathbf{C}^{-1}+\mathbf{D}^{-1}\right) \mathbf{\rho }=e%
\mathbf{\mathbf{\bar{C}}}^{-1}\mathbf{\rho .}
\end{equation}%
For convenience, we use effective capacitance matrix $\mathbf{\bar{C}}$
defined as $\mathbf{\bar{C}}^{-1}=\mathbf{C}^{-1}+\mathbf{D}^{-1}/e$. Its
matrix elements can be written as%
\begin{equation}
\mathbf{\bar{C}=}\left( 
\begin{array}{cc}
\bar{C}_{\mathrm{1}}+\bar{C}_{\mathrm{x}} & -\bar{C}_{\mathrm{x}} \\ 
-\bar{C}_{\mathrm{x}} & \bar{C}_{\mathrm{2}}+\bar{C}_{\mathrm{x}}%
\end{array}%
\right) ,
\end{equation}%
where $\bar{C}_{\mathrm{1}}$, $\bar{C}_{\mathrm{2}}$, and $\bar{C}_{\mathrm{x%
}}$ approach $C_{\mathrm{1}}$, $C_{\mathrm{2}}$, and $C_{\mathrm{x}}$,
respectively, of Eq. (\ref{EqrCV}) in the limit of large $D_{c}$ ($\gg C_{c}$%
). For a soft edge potential with large $D_{c}$, such as in AlGaAs/GaAs
heterostructures ($\mathbf{C}^{-1}\gg \mathbf{D}^{-1}/e$), $\mathbf{\bar{C}}$
can be approximated to the geometric capacitance matrix $\mathbf{C}$.
Nevertheless, we shall use $\mathbf{\bar{C}}$ in the following analysis.

The difference in the electrochemical potentials induces a tunneling current 
$I_{\mathrm{x}}=g\left( \mu _{2}-\mu _{1}\right) /e$ between the channels.
The resulting current $I_{\mathrm{c}}$ on channel $c$ is determined by the
charge conservation ($\partial \rho _{1}/\partial t=-\partial I_{1}/\partial
x+I_{\mathrm{x}}$, $\partial \rho _{2}/\partial t=-\partial I_{2}/\partial
x-I_{\mathrm{x}}$), which can be written as%
\begin{equation}
\frac{\partial }{\partial t}\mathbf{\rho =-}\frac{\partial }{\partial x}%
\mathbf{I-g\mu }/e
\end{equation}%
with current $\mathbf{I}=\left( 
\begin{array}{c}
I_{1} \\ 
I_{2}%
\end{array}%
\right) $ and tunneling conductance $\mathbf{g}=g\left( 
\begin{array}{cc}
1 & -1 \\ 
-1 & 1%
\end{array}%
\right) $. As the current is related to the electrochemical potential ($%
\mathbf{I}=\mathbf{\sigma \mu }/e$) with the conductance matrix $\mathbf{%
\sigma }=\left( 
\begin{array}{cc}
\sigma _{1} & 0 \\ 
0 & \sigma _{2}%
\end{array}%
\right) $, the differential equation can be written in terms of $\mathbf{I}$
as%
\begin{equation}
\frac{\partial }{\partial t}\mathbf{I=-\sigma \bar{C}}^{-1}\frac{\partial }{%
\partial x}\mathbf{I-\sigma \bar{C}}^{-1}\mathbf{g\sigma }^{-1}\mathbf{I.}
\label{dIdt-EOM}
\end{equation}%
This can be used to analyze the dynamics of the current $\mathbf{I}\left(
x,t\right) $ as well as the transport eigenmodes of the system. Once $%
\mathbf{I}$ is determined, one can translate this to the electrochemical
potential $\mathbf{\mu }=e\mathbf{\sigma }^{-1}\mathbf{I}$ and the charge
density $\mathbf{\rho }=\mathbf{\bar{C}\sigma }^{-1}\mathbf{I}$.

Before solving Eq. (\ref{dIdt-EOM}), we describe the relation to the
standard field theory on one-dimensional channels. Generally, the Lagrangian
density can be written as%
\begin{equation}
L=\frac{\hbar }{4\pi }\sum_{i,j}\left( K_{ij}\partial _{t}\phi _{i}\partial
_{x}\phi _{j}+V_{ij}\partial _{x}\phi _{i}\partial _{x}\phi _{j}\right)
\end{equation}%
for a boson field $\phi _{i}$ on channel $i$, where $\mathbf{K}=\left(
K_{ij}\right) $ is an integer-valued symmetric matrix and $\mathbf{V}=\left(
V_{ij}\right) $ is a positive definite matrix \cite{BookWen,Wen-PRL1990}. By
using the Euler-Lagrange equation $\partial L/\partial \phi _{i}=\partial
_{\mu }\left[ \partial L/\partial \left( \partial _{\mu }\phi _{i}\right) %
\right] $ with variable $\mu =\left\{ x,t\right\} $, one can derive an
equation of motion%
\begin{equation}
\sum_{j}K_{ij}\partial _{t}\left( \partial _{x}\phi _{j}\right)
=-\sum_{j}V_{ij}\partial _{x}\left( \partial _{x}\phi _{j}\right) .
\end{equation}%
As the charge density $\rho _{i}=-\frac{e}{2\pi }\partial _{x}\phi _{j}$ is
related to the field, this can be transformed to

\begin{equation}
\mathbf{K}\frac{\partial }{\partial t}\mathbf{\rho =-V}\frac{\partial }{%
\partial x}\mathbf{\rho ,}
\end{equation}%
which is equivalent to Eq. (\ref{dIdt-EOM}) until the scattering (second)
term. This comparison suggests that the relations $\mathbf{K}=\sigma _{q}%
\mathbf{\sigma }^{-1}$ and $\mathbf{V}=\sigma _{q}\mathbf{\bar{C}}^{-1}$
exist between the parameters in both models. For instance, velocity
parameters $v_{1}$, $v_{2}$, and $v_{12}$, used in Kane et al.'s seminal
paper \cite{KaneFisherPolchinski-PRL1994} for the $\nu =2/3$ case, can be
described with elements of $\mathbf{\bar{C}}$ as 
\begin{eqnarray}
v_{1} &=&\frac{\sigma _{q}\left( \bar{C}_{2}+\bar{C}_{\mathrm{x}}\right) }{%
\bar{C}_{1}\bar{C}_{2}+\left( \bar{C}_{1}+\bar{C}_{2}\right) \bar{C}_{%
\mathrm{x}}} \\
v_{2} &=&\frac{\sigma _{q}\left( \bar{C}_{1}+\bar{C}_{\mathrm{x}}\right) }{%
\bar{C}_{1}\bar{C}_{2}+\left( \bar{C}_{1}+\bar{C}_{2}\right) \bar{C}_{%
\mathrm{x}}} \\
v_{12} &=&\frac{\sigma _{q}\bar{C}_{\mathrm{x}}}{\bar{C}_{1}\bar{C}%
_{2}+\left( \bar{C}_{1}+\bar{C}_{2}\right) \bar{C}_{\mathrm{x}}}.
\end{eqnarray}%
With these relations, the parameters in the effective theory can be related
to realistic parameters: the capacitances of the channel geometries and the
density of states of the channels.

The scattering term, the second term of Eq. (\ref{dIdt-EOM}), has been
obtained in the incoherent tunneling regime, where local potential $\mu
_{c}\left( x\right) $ is well defined by assuming local equilibrium between
successive scattering events \cite{Nosiglia-PRB2018}. This has been used to
calculate shot noise characteristics and heat transport in fractional
quantum Hall systems \cite{Spanslatt-PRL2019,Park-PRB2019}. Here, we apply
Eq. (\ref{dIdt-EOM}) with this scattering term to investigate time-dependent
dynamics in the incoherent regime.

\section{Transport eigenmodes}

We shall derive the transport eigenmodes of Eq. (\ref{dIdt-EOM}) at angular
frequency $\omega $. As the scattering term is dissipative, the
corresponding momentum $k$ generally takes a complex value. By assuming
solutions in a form of $\mathbf{I}\left( x,t\right) =\mathbf{\tilde{I}}\exp %
\left[ i\left( kx-\omega t\right) \right] $ with eigenmode $\mathbf{\tilde{I}%
}$, Eq. (\ref{dIdt-EOM}) is transformed to%
\begin{equation}
k\mathbf{\tilde{I}}=\left( \omega \mathbf{\bar{C}\sigma }^{-1}+i\mathbf{%
g\sigma }^{-1}\right) \mathbf{\tilde{I}}=\mathbf{M\tilde{I},}
\end{equation}%
from which $\mathbf{\tilde{I}}$ and $k$ can be obtained by diagonalizing $%
\mathbf{M}=\left( \omega \mathbf{\bar{C}}+i\mathbf{g}\right) \mathbf{\sigma }%
^{-1}$.

It is convenient to introduce dimensionless parameters 
\begin{eqnarray}
\xi  &=&\frac{\omega \bar{C}_{\mathrm{x}}+ig}{\omega \left( \bar{C}_{1}-m%
\bar{C}_{2}\right) } \\
m &=&\sigma _{1}/\sigma _{2}
\end{eqnarray}%
to characterize the modes, where $m=1$ for the $\nu =2$ case and $m=-3$ for
the $\nu =2/3$ case. The complex value $\xi $ determines the relative
strength of the interaction ($\bar{C}_{\mathrm{x}}$) in the real part and of
the tunneling ($g$) in the imaginary part. The two eigenmodes (labeled by I
and II) are written as%
\begin{eqnarray}
\mathbf{\tilde{I}}_{\mathrm{I}} &=&\binom{1}{\tilde{I}_{\mathrm{I}}}\text{
with }\tilde{I}_{\mathrm{I}}=\frac{-2\xi }{1+\left( 1-m\right) \xi
+X_{m}\left( \xi \right) }  \label{I1I2} \\
\mathbf{\tilde{I}}_{\mathrm{II}} &=&\binom{\tilde{I}_{\mathrm{II}}}{1}\text{
with }\tilde{I}_{\mathrm{II}}=\frac{1+\left( 1-m\right) \xi -X_{m}\left( \xi
\right) }{-2\xi }
\end{eqnarray}%
where the $\xi $ dependent function $X_{m}\left( \xi \right) $ is defined as%
\begin{equation}
X_{m}\left( \xi \right) =\sqrt{1+2\left( 1-m\right) \xi +\left( 1+m\right)
^{2}\xi ^{2}}.
\end{equation}%
The corresponding $k$ values are obtained as

\begin{eqnarray}
k_{\mathrm{I}} &=&\frac{\omega }{2\sigma _{1}}\{\bar{C}_{1}\left[ 1+\left(
1+m\right) \xi +X_{m}\left( \xi \right) \right]   \label{kIkII} \\
&&+m\bar{C}_{2}\left[ 1-\left( 1+m\right) \xi -X_{m}\left( \xi \right) %
\right] \}  \notag \\
k_{\mathrm{II}} &=&\frac{\omega }{2\sigma _{1}}\{\bar{C}_{1}\left[ 1+\left(
1+m\right) \xi -X_{m}\left( \xi \right) \right]  \\
&&+m\bar{C}_{2}\left[ 1-\left( 1+m\right) \xi +X_{m}\left( \xi \right) %
\right] \}  \notag
\end{eqnarray}%
for mode I and II, respectively.

In this way, the eigenmodes, $\mathbf{\tilde{I}}_{\mathrm{I}}$ and $\mathbf{%
\tilde{I}}_{\mathrm{II}}$, can be written with a single complex parameter $%
\xi =\xi ^{\prime }+i\xi ^{\prime \prime }$ with the real part $\xi ^{\prime
}$ and the imaginary part $\xi ^{\prime \prime }$. Mode I and II approach
the isolated channels 1 and 2, respectively, in the limit of $\xi
\longrightarrow 0$. In the limit of either $\xi \longrightarrow \infty $ or $%
\xi \longrightarrow i\infty $, the eigenmodes approach the pure neutral mode 
$\binom{1}{-1}$ with vanishing total current and the pure charge mode $%
\binom{m}{1}$ with equal electrochemical potential $\mathbf{\mu }\propto 
\binom{1}{1}$.

\begin{figure*}[tbp]
\begin{center}
\includegraphics[width = 6.6 in]{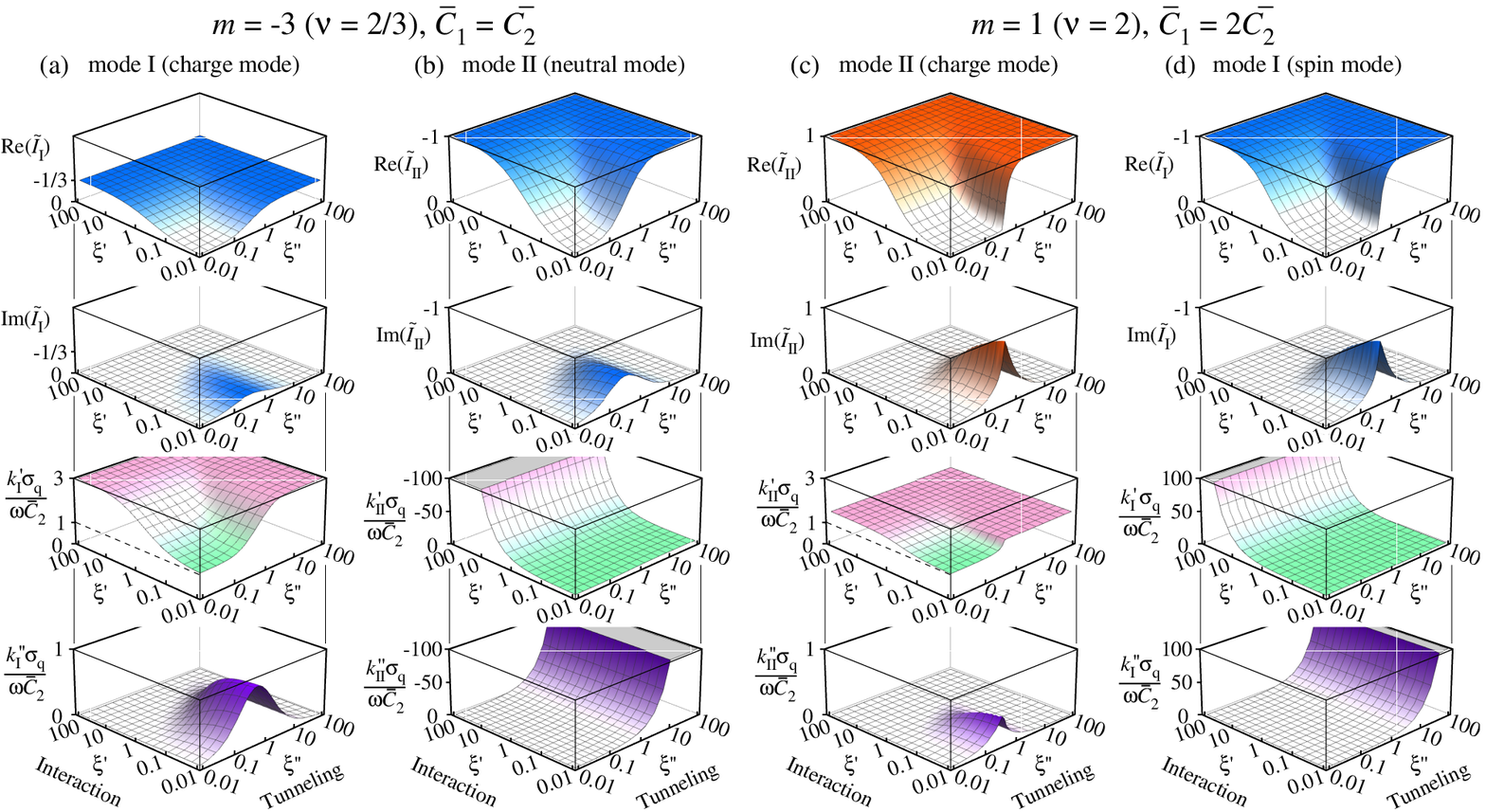}
\end{center}
\caption{(a and b) Plasmon eigenmodes for the $\protect\nu =2/3$ case ($m=-3$%
) with $\bar{C}_{1}=\bar{C}_{2}$. Mode I in (a) and mode II in (b) show the
(quasi-) charge mode and the (quasi-) neutral mode, respectively. (c and d)
Plasmon eigenmodes for the $\protect\nu =2$ case ($m=1$) with $\bar{C}_{1}=2%
\bar{C}_{2}$. Mode II in (c) and mode I in (d) show the (quasi-) charge mode
and the (quasi-) spin mode, respectively. Each system can be characterized
by dimensionless parameters, $\protect\xi ^{\prime }$ for the interaction
and $\protect\xi ^{\prime \prime }$ for the tunneling. The topmost and the
second top panels show real and imaginary parts, respectively, of the
eigenmodes $\tilde{I}_{\mathrm{I}}$ and $\tilde{I}_{\mathrm{II}}$ (blue for
negative and red for positive). The eigenmodes reach uncoupled modes ($%
\tilde{I}_{\mathrm{I}}=\tilde{I}_{\mathrm{II}}=0$) at the limit of $\protect%
\xi ^{\prime }\rightarrow 0$ and $\protect\xi ^{\prime \prime }\rightarrow 0$%
, and the pure charge and neutral modes at the limit of $\protect\xi %
^{\prime }\rightarrow \infty $ or $\protect\xi ^{\prime \prime }\rightarrow
\infty $. The third and bottom panels show $k^{\prime }\protect\sigma _{%
\mathrm{q}}/\protect\omega \bar{C}_{2}$ for wavenumber $k^{\prime }$ (green
for long and pink for short wavelength)\ and $k^{\prime \prime }\protect%
\sigma _{\mathrm{q}}/\protect\omega \bar{C}_{2}$ for decay rate $k^{\prime
\prime }$ (purple for strong dissipation), respectively. Positive $k^{\prime
}$ and $k^{\prime \prime }$ in (a), (c), and (d) represent the right movers,
while negative values in (b) indicate the left-mover.}
\end{figure*}

Figure 2 shows the eigenmode, the real parts $\mathrm{Re}\left( \tilde{I}%
\right) $ in the top panel and the imaginary parts $\mathrm{Im}\left( \tilde{%
I}\right) $ in the second top panel, as well as the normalized $k=k^{\prime
}+ik^{\prime \prime }$ value, $k^{\prime }\sigma _{\mathrm{q}}/\omega \bar{C}%
_{2}$ in the third panel and $k^{\prime \prime }\sigma _{\mathrm{q}}/\omega 
\bar{C}_{2}$ in the bottom panel. The negative $k^{\prime }$ and $k^{\prime
\prime }$ values in Fig. 2(b) correspond to left-moving excitations. For
each panel, the interaction parameter $\xi ^{\prime }$ and the tunneling
parameter $\xi ^{\prime \prime }$ are varied across a wide range of 0.01 -
100. While $\xi ^{\prime }=\bar{C}_{\mathrm{x}}/\left( \bar{C}_{1}+3\bar{C}%
_{2}\right) $ is determined by the ratio of the capacitances, $\xi ^{\prime
\prime }=g/\omega \left( \bar{C}_{1}-m\bar{C}_{2}\right) $ depends on the
frequency $\omega $. The dc measurement corresponds to $\xi ^{\prime \prime
}\rightarrow \infty $ unless the scattering is absent ($g=0$).

The eigenmodes for the $\nu =2/3$ case ($m=-3$) are shown in Figs. 2(a) and
2(b). The top panels show that the pure charge mode with $\tilde{I}_{\mathrm{%
I}}=-1/3$ and the pure neutral mode with $\tilde{I}_{\mathrm{II}}=-1$ appear
when either $\xi ^{\prime }$ or $\xi ^{\prime \prime }$ becomes much greater
than 1. The roles of the interaction and the tunneling appear differently in
the $k$ values of the neutral mode in Fig. 2(b). While the $|k^{\prime }|$\
value in the third panel increases (or the velocity decreases) with
increasing $\xi ^{\prime }$, the $|k^{\prime \prime }|$ value in the bottom
panel increases (the dissipation increases) with increasing $\xi ^{\prime
\prime }$. In contrast, the charge mode in Fig. 2(a) shows mild dissipation (%
$k^{\prime \prime }<k^{\prime }$) in the narrow parameter range around $\xi
^{\prime \prime }\sim 1$ for $\xi ^{\prime }\lesssim 1$, and small or
negligible dissipation ($k^{\prime \prime }\ll k^{\prime }$) in the
low-frequency or strong tunneling regime ($\xi ^{\prime \prime }>1$). In
this way, the tunneling stabilizes the charge mode by degrading the neutral
mode.

Quite similar characteristics are seen for the $\nu =2$ case ($m=1$) in
Figs. 2(c) and 2(d). The top panels show that the pure charge mode with $%
\tilde{I}_{\mathrm{II}}=1$ in Fig. 2(c) and the pure spin mode with $\tilde{I%
}_{\mathrm{I}}=-1$ in Fig. 2(d) at either $\xi ^{\prime }\gg 1$ or $\xi
^{\prime \prime }\gg 1$. The bottom panels show that the tunneling induces
stable charge transport ($k^{\prime \prime }<k^{\prime }$) in Fig. 2(c) and
dissipative spin transport ($k^{\prime \prime }\gg k^{\prime }$) in Fig.
2(d). Therefore, the same features can be studied in both $\nu =2/3$ and $2$
cases within the plasmon scattering approach.

\section{Transmission \& reflection measurement}

\begin{figure}[tbp]
\begin{center}
\includegraphics[width = 2.8 in]{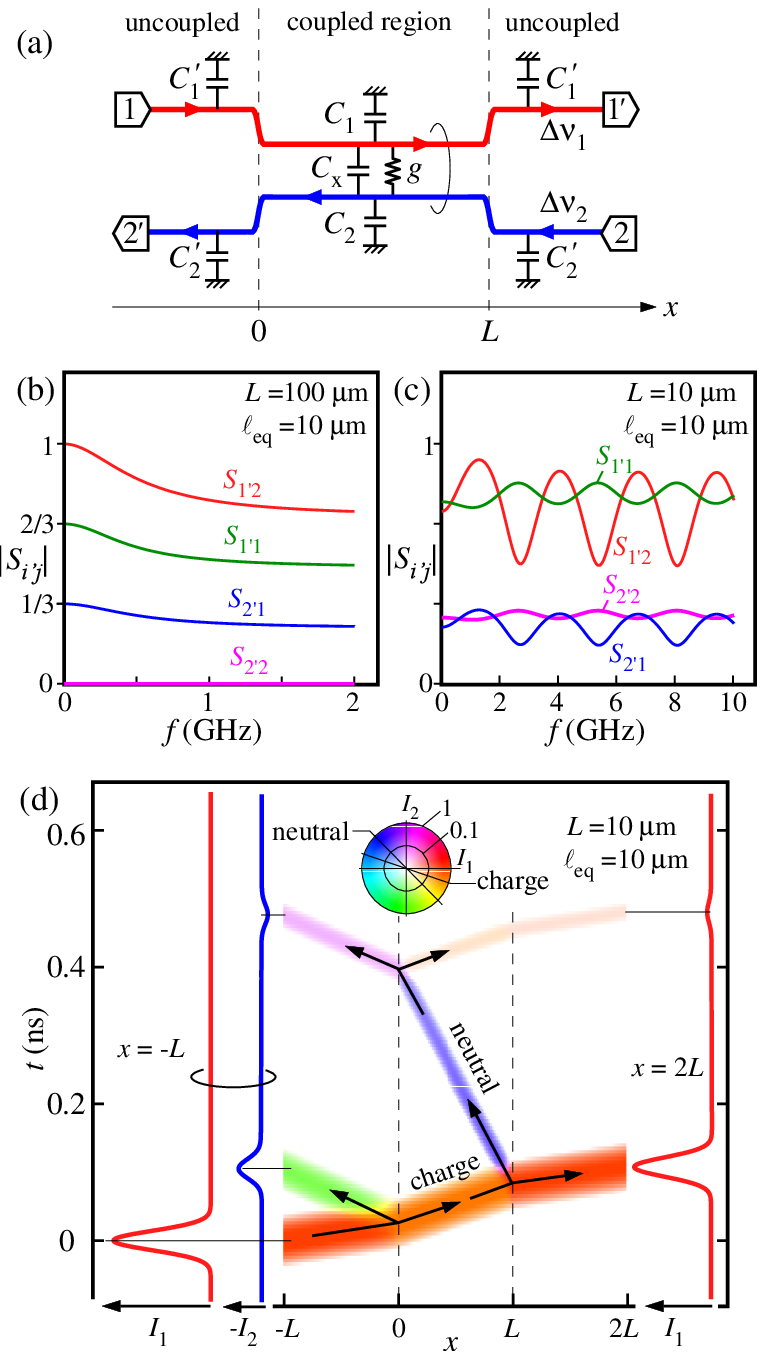}
\end{center}
\caption{(a) Possible experimental setup for transmission and reflection
measurements through the coupled region of length $L$. Plots in (b)-(d) are
obtained with $\Delta \protect\nu _{1}=1$, $\Delta \protect\nu _{2}=-1/3$, $%
\bar{C}_{1}=\bar{C}_{2}=\bar{C}_{1}^{\prime }=\bar{C}_{2}^{\prime }=$ 0.1
nF/m, $\bar{C}_{\mathrm{x}}=$ 0.4 nF/m, and $\ell _{\mathrm{eq}}=$ 10 $%
\protect\mu $m (corresponding to $g=$ 1.9 S/m). (b) and (c) Magnitude of the
scattering matrix element $\left\vert S_{i^{\prime }j}\right\vert $ as a
function of frequency $f$ for a long channel with $L$ = 100 $\protect\mu $m
in (b) and short channel with $L$ = 10 $\protect\mu $m in (c). (d) Space ($x$%
) - time ($t$) evolution of the current profiles $I_{1}\left( x,t\right) $
and $I_{2}\left( x,t\right) $. A current pulse $I_{1}\left( t\right) $ (the
leftmost trace) is injected at $x=-L$, and resulting currents, $I_{1}\left(
x,t\right) $ and $I_{2}\left( x,t\right) $, are shown. In the central color
plot, a specific color scale is used: orange for the charge mode, blue for
the neutral mode, red for the uncoupled channel 1, and green or purple for
the uncoupled channel 2. Transmitted current $I_{1}\left( t\right) $ at $x=2L
$ (the rightmost trace) and reflected current$-I_{2}\left( t\right) $ at $%
x=-L$ (the second-left trace) also are shown.}
\end{figure}

The transport eigenmodes can be studied experimentally by investigating
transmission and reflection characteristics of an interaction region with
length $L$, as shown in Fig. 3(a) \cite{Bocquillon-NatCom2013}. Here, we
focus on the fractional case of $\nu =2/3$, while similar results are
expected for other cases. Current can be excited from ports 1 or 2, and the
output current can be measured at ports 1$^{\prime }$ and 2$^{\prime }$. The
intrachannel interaction in the uncoupled regions is parametrized with
effective capacitances $\bar{C}_{1}^{\prime }$ and $\bar{C}_{2}^{\prime }$
(geometric capacitance $C_{c}^{\prime }$ and the density of states). The
scattering matrix element $S_{i^{\prime }j}$ from input port $j$ to output
port $i^{\prime }$ can be derived by using the eigenmodes in the interacting
region ($\tilde{I}_{\mathrm{I}}$, $\tilde{I}_{\mathrm{II}}$, $k_{\mathrm{I}}$%
, and $k_{\mathrm{II}}$ derived in the previous section) and current
conservation at the boundaries. By using the complex phases $\phi _{\mathrm{I%
}}=k_{\mathrm{I}}L$ and $\phi _{\mathrm{II}}=k_{\mathrm{II}}L$ acquired in
the interacting region, the scattering matrix is written as 
\begin{eqnarray}
\mathbf{S} &=&\frac{1}{1-\text{ }\tilde{I}_{\mathrm{I}}\text{ }\tilde{I}_{%
\mathrm{II}}e^{i\phi _{\mathrm{I}}}e^{-i\phi _{\mathrm{II}}}}\times 
\label{Smat} \\
&&\left( 
\begin{array}{cc}
e^{i\phi _{\mathrm{I}}}\left( 1-\tilde{I}_{\mathrm{I}}\text{ }\tilde{I}_{%
\mathrm{II}}\right) \text{ } & \tilde{I}_{\mathrm{II}}\left( 1-e^{i\phi _{%
\mathrm{I}}}e^{-i\phi _{\mathrm{II}}}\right)  \\ 
\tilde{I}_{\mathrm{I}}\left( 1-e^{i\phi _{\mathrm{I}}}e^{-i\phi _{\mathrm{II}%
}}\right)  & e^{-i\phi _{\mathrm{II}}}\left( 1-\tilde{I}_{\mathrm{I}}\text{ }%
\tilde{I}_{\mathrm{II}}\right) \text{ }%
\end{array}%
\right) .  \notag
\end{eqnarray}

First, we consider the dc characteristics with this $S$ matrix. For the $\nu
=2/3$ case ($m=-3$), at the dc limit ($\omega \rightarrow 0$), eigenmodes
with the characteristics of Eqs. (\ref{I1I2}) and (\ref{kIkII}) are well
approximated to the pure charge mode with $\tilde{I}_{\mathrm{I}}=-1/3$ and $%
k_{\mathrm{I}}\simeq 0$, and the pure neutral mode with $\tilde{I}_{\mathrm{%
II}}=-1$ and $k_{\mathrm{II}}\simeq -2i\frac{g}{\sigma _{1}}$. The so-called
equilibration length $\ell _{\mathrm{eq}}$, defined as the decay length of
the neutral mode ($\left\vert e^{-ik_{\mathrm{II}}\ell _{\mathrm{eq}%
}}\right\vert =1/e$) \cite{Grivnin-PRL2014,CJLin-PRB2019}, is related to $g$
as 
\begin{equation}
\ell _{\mathrm{eq}}=\frac{\sigma _{1}}{2g}.
\end{equation}%
For a long channel much greater than $\ell _{\mathrm{eq}}$, the $S$ matrix
at the dc limit is reduced to 
\begin{equation}
\mathbf{S}_{\mathrm{dc}}\simeq \left( 
\begin{array}{cc}
2/3 & -1 \\ 
\text{ }-1/3 & 0%
\end{array}%
\right) ,  \label{SmatDC}
\end{equation}%
which indicates the current partition from port 1 to port 1$^{\prime }$ with
the transmission factor 2/3 and to port 2$^{\prime }$ with the reflection
factor 1/3 ($-S_{2^{\prime }1}$ as the polarity is defined as positive for a
right-moving current).

For high-frequency transport with finite $\xi ^{\prime \prime }$, mode II
represents the quasi-neutral mode propagating to the left, and $k_{\mathrm{II%
}}$ has a significantly large negative imaginary part. By using a
sufficiently long channel with $\left\vert e^{-i\phi _{\mathrm{II}%
}}\right\vert \ll 1$, Eq. (\ref{Smat}) can be approximated to%
\begin{equation}
\mathbf{S}\simeq \left( 
\begin{array}{cc}
e^{i\phi _{\mathrm{I}}}\left( 1-\tilde{I}_{\mathrm{I}}\text{ }\tilde{I}_{%
\mathrm{II}}\right) & \text{ }\tilde{I}_{\mathrm{II}} \\ 
\text{ }\tilde{I}_{\mathrm{I}} & e^{-i\phi _{\mathrm{II}}}\left( 1-\tilde{I}%
_{\mathrm{I}}\text{ }\tilde{I}_{\mathrm{II}}\right)%
\end{array}%
\right) .  \label{SmatApprox}
\end{equation}%
At this limit, $\tilde{I}_{\mathrm{I}}$ and $\tilde{I}_{\mathrm{II}}$ can be
determined directly from $S_{2^{\prime }1}$ and $S_{1^{\prime }2}$,
respectively, and $k_{\mathrm{I}}$ and $k_{\mathrm{II}}$ can be determined
from the phase factor in $S_{1^{\prime }1}$ and $S_{2^{\prime }2}$,
respectively. Figure 3(b) shows $\left\vert S_{i^{\prime }j}\right\vert $ as
a function of frequency $f$ for channel length $L$ = 100 $\mu $m ($\gg $ $%
\ell _{\mathrm{eq}}$ = 10 $\mu $m), where the eigenmodes at higher frequency
deviate from the values at the dc limit. In contrast, when the channel
length is much shorter with $\left\vert e^{i\phi _{\mathrm{I}}}e^{-i\phi _{%
\mathrm{II}}}\right\vert \sim 1$, the scattering matrix elements in Eq. (\ref%
{Smat}) oscillate with the factor $e^{i\phi _{\mathrm{I}}}e^{-i\phi _{%
\mathrm{II}}}$ that arises from the plasmon interference of the two modes.
Such interference can be seen in Fig. 3(c), where the channel length $L$ =
10 $\mu $m is set equal to $\ell _{\mathrm{eq}}$. Observation of the
interference pattern can be used to identify all parameters in the plasmon
model.

\section{Time-domain experiment}

One can investigate the eigenmodes by using time-domain experiments \cite%
{Kamata-NatNano2014,Hashisaka-NatPhys2017,CJLin}. For example, by using the
same device structure as that shown in Fig. 3(a), a sharp current pulse can
be introduced to port 1 and the resulting currents at ports 1' and 2' can be
measured with time-resolved detectors. Such an experiment is simulated in
Fig. 3(d), where non-interacting channels ($-L<x<0$ and $L<x<2L$) are
attached to the interacting region of length $L$ ($0<x<L$). By applying a
current pulse in the form of $I_{1}(t)$ at $x=-L$ as shown in the leftmost
trace, the current distributions $I_{1}(x,t)$ and $I_{2}(x,t)$ can be
calculated using Eq. (\ref{dIdt-EOM}). Note that the calculus of finite
differences in space and time is problematic for this equation. Instead, we
considered a periodic pulse with a long period, and the scattering matrix of
Eq. (\ref{Smat}) is used for each frequency component in the Fourier
spectrum of the incident wave. The space-time solution is obtained by using
the inverse Fourier transform.

The current distributions are plotted in a specific color scale with hue for
the mode $\left( I_{1},I_{2}\right) $ and brightness for the magnitude in a
logarithmic scale. One can see that the incident charge packet is
fractionalized into different modes (colors) at $x=0$ and $x=L$. The model
predicts that the transmitted current $I_{1}(t)$ at $x=2L$ (the rightmost
trace) and the reflected current $-I_{2}(t)$ at $x=-L$ (the second trace
from the left) should be obtained experimentally. One can investigate the
transport eigenmodes by analyzing the amplitude and the time-of-flight of
the fractionalized wave packets.

\section{Summary}

We have provided a plasmon model for two chiral edge channels in the
presence of Coulomb interaction and disorder-induced tunneling by using a
formulation in the incoherent regime. This can be used to systematically
understand the transport characteristics of various cases, including
copropagating integer channels at $\nu =2$ and counterpropagating
integer/fractional channels at $\nu =2/3$. We have proposed two experimental
schemes. These would use frequency- and time-domain measurements to
investigate the transmission and reflection characteristics of a coupling
region with a finite length. Particularly, when the coupling length is
comparable to or shorter than the equilibration length, rich characteristics
associated with the fractionalization at the boundaries are expected. While
representative cases with $\nu =2$ and 2/3 are considered in this study, the
model can be extended to more complicated quantum Hall channels (more than
two channels) and other topological systems with multiple channels \cite%
{TopoQC-Nayak08,Polkovnikov-RMP2011,BookBernevig}. This may be useful in
identifying an appropriate and effective model for edge reconstructions from
several candidates \cite{Beenakker-PRL90,Meir-PRL94}.

\begin{acknowledgments}
We thank Masayuki Hashisaka, Tokuro Hata, Koji Muraki, and Yasuhiro Tokura
for fruitful discussions. This work was supported by JSPS KAKENHI
(JP15H05854, JP19H05603).
\end{acknowledgments}


\begin{thebibliography}{99}
\bibitem{BookEzawa} Z. F. Ezawa, Quantum Hall Effects: Recent Theoretical
and Experimental Developments, 3rd edition, (World Scientific, Singapore,
2013).

\bibitem{BookWen} X-G Wen, Quantum Field Theory of Many-body Systems (Oxford
Univ. Press, 2004).

\bibitem{BookGiamarchi} T. Giamarchi, Quantum Physics in One Dimension
(Oxford Univ. Press, 2004).

\bibitem{Bocquillon-NatCom2013} E. Bocquillon et al., Separation of neutral
and charge modes in one-dimensional chiral edge channels. Nat. Commun. 4,
1839 (2013).

\bibitem{Inoue-PRL2014} H. Inoue, A. Grivnin, N. Ofek, I. Neder, M. Heiblum,
V. Umansky, and D. Mahalu, Charge fractionalization in the integer quantum
Hall effect, Phys. Rev. Lett. 112, 166801 (2014).

\bibitem{Hashisaka-NatPhys2017} M. Hashisaka, N. Hiyama, T. Akiho, K.
Muraki, and T. Fujisawa, Waveform measurement of charge- and spin-density
wavepackets in a chiral Tomonaga-Luttinger liquid, Nat. Phys. 13, 559 (2017).

\bibitem{Bid-PRL2009} A. Bid, N. Ofek, M. Heiblum, V. Umansky, and D.
Mahalu, Shot noise and charge at the 2/3 composite fractional quantum Hall
state, Phys. Rev. Lett. 103, 236802 (2009).

\bibitem{Sabo-NatPhys17} R. Sabo, I. Gurman, A. Rosenblatt, F. Lafont, D.
Banitt, J. Park, M. Heiblum, Y. Gefen, V. Umansky, D. Mahalu, Edge
reconstruction in fractional quantum Hall~states, Nat. Phys. 13, 491 (2017).

\bibitem{Lafont-Science2019} F. Lafont, A. Rosenblatt, M. Heiblum, V.
Umansky, Counter-propagating charge transport in the quantum Hall effect
regime, Science 363, 54 (2019).

\bibitem{Kamata-PRB10} H. Kamata, T. Ota, K. Muraki, and T. Fujisawa,
Voltage-controlled group velocity of edge magnetoplasmon in the quantum Hall
regime, Phys. Rev. B 81, 085329 (2010).

\bibitem{Altimiras-NatPhys10} C. Altimiras, H. le Sueur, U. Gennser, A.
Cavanna, D. Mailly, and F. Pierre, Non-equilibrium edge-channel spectroscopy
in the integer quantum Hall regime. Nat. Phys. 6, 34 (2010).

\bibitem{Kamata-NatNano2014} H. Kamata, N. Kumada, M. Hashisaka, K. Muraki,
and T. Fujisawa, Fractionalized wave packets from an artificial
Tomonaga-Luttinger liquid. Nat. Nanotech. 9, 177 (2014).

\bibitem{Itoh-PRL18} K. Itoh, R. Nakazawa, T. Ota, M. Hashisaka, K. Muraki
and T. Fujisawa, Signatures of a Nonthermal Metastable State in
Copropagating Quantum Hall Edge Channels. Phys. Rev. Lett. 120, 197701
(2018).

\bibitem{Ji-Nature03} Y. Ji, Y. C. Chung, D. Sprinzak, M. Heiblum, D.
Mahalu, and H. Shtrikman, An electronic Mach-Zehnder interferometer, Nature
422, 415 (2003).

\bibitem{Berg-PRL2009} E. Berg, Y. Oreg, E. A. Kim, and F. von Oppen,
Fractional charges on an integer quantum Hall edge. Phys. Rev. Lett. 102,
236402 (2009).

\bibitem{KaneFisher-PRB1995} C. L. Kane and M. P. A. Fisher, Contacts and
edge-state equilibration in the fractional quantum Hall effect. Phys. Rev. B
52, 17393 (1995).

\bibitem{KaneFisherPolchinski-PRL1994} C. L. Kane, M. P. A. Fisher, and J.
Polchinski, Randomness at the Edge: Theory of quantum Hall transport at
filling $\nu $ = 2/3, Phys. Rev. Lett. 72, 4129 (1994).

\bibitem{Grivnin-PRL2014} A. Grivnin, H. Inoue, Y. Ronen, Y. Baum, M.
Heiblum, V. Umansky, and D. Mahalu, Non-equilibrated counter propagating
edge modes in the fractional quantum Hall regime, Phys. Rev. Lett. 113,
266803 (2014)

\bibitem{CJLin-PRB2019} C. J. Lin, R. Eguchi, M. Hashisaka, T. Akiho, K.
Muraki, and T. Fujisawa, Charge equilibration in integer and fractional
quantum Hall edge channels in a generalized Hall-bar device, Phys. Rev. B
99, 195304 (2019).

\bibitem{CJLin} C. J. Lin, M. Hashisaka, T. Akiho, K. Muraki, and T.
Fujisawa, Quantized charge fractionalization at quantum Hall Y junctions in
the disorder dominated regime, Nature Commun. 12, 131 (2021).

\bibitem{Safi-PRB1995} I. Safi, and H. J. Schulz, Transport in an
inhomogeneous interacting one-dimensional system, Phys. Rev. B 52, R17040(R)
(1995).

\bibitem{SafiEurPhys} I. Safi, A dynamic scattering approach for a gated
interacting wire, Eur. Phys. J. B 12, 451 (1999).

\bibitem{Protopopov-AnnPhys2017} I. V. Protopopov, T. Gefen, and A. D.
Mirlin, Transport in a disordered $\nu $ = 2/3 fractional quantum Hall
junction, Ann. Phys. 385, 287 (2017).

\bibitem{Hashisaka-PRB2012} M. Hashisaka, K. Washio, H. Kamata, K. Muraki,
and T. Fujisawa, Distributed electrochemical capacitance evidenced in
high-frequency admittance measurements on a quantum Hall device, Phys. Rev.
B 85, 155424 (2012).

\bibitem{MacDonald-PRL1990} A. H. MacDonald, Edge states in the
fractional-quantum-Hall effect regime, Phys. Rev. Lett. 64, 220 (1990).

\bibitem{Wen-PRL1990} X. G. Wen, Electrodynamical properties of gapless edge
excitations in the fractional quantum Hall states, Phys. Rev. Lett. 64, 2206
(1990).

\bibitem{Nosiglia-PRB2018} C. Nosiglia, J. Park, B. Rosenow, and Y. Gefen,
Incoherent transport on the $\nu $ = 2/3 quantum Hall edge, Phys. Rev. B 98,
115408 (2018).

\bibitem{Hashisaka-RevPhys2018} M. Hashisaka and T. Fujisawa,
Tomonaga-Luttinger-liquid nature of edge excitations in integer quantum Hall
edge channels, Reviews in Physics 3, 32 (2018).

\bibitem{Gabelli-PRL2007} J. Gabelli, G. Feve, T. Kontos, J. M. Berroir, B.
Placais, D. C. Glattli, B. Etienne, Y. Jin, and M. Buttiker, Relaxation Time
of a Chiral Quantum R-L Circuit, Phys. Rev. Lett. 98, 166806 (2007).

\bibitem{Spanslatt-PRL2019} C. Sp\aa nsl\"{a}tt, J. Park, Y. Gefen, A. D.
Mirlin, Topological Classification of Shot Noise on Fractional Quantum Hall
Edges, Phys. Rev. Lett 123, 137701 (2019).

\bibitem{Park-PRB2019} J. Park, A. D. Mirlin, B. Rosenow, and Y. Gefen,
Noise on complex quantum Hall edges: Chiral anomaly and heat diffusion,
Phys. Rev. B 99, 161302(R) (2019).

\bibitem{TopoQC-Nayak08} C. Nayak, S. H. Simon, A. Stern, M. Freedman and S.
Das Sarma, Non-Abelian anyons and topological quantum computation, Rev. Mod.
Phys. \textbf{80}, 1083 (2008).

\bibitem{Polkovnikov-RMP2011} A. Polkovnikov, K. Sengupta, A. Silva, and M.
Vengalattore, Nonequilibrium dynamics of closed interacting quantum systems,
Rev. Mod. Phys. 83, 863 (2011).

\bibitem{BookBernevig} B. A. Bernevig, Topological Insulators and
Topological Superconductors, (Princeton Univ. Press 2013).

\bibitem{Beenakker-PRL90} C. W. J. Beenakker, Edge channels for the
fractional quantum Hall effect. Phys. Rev. Lett. 64, 216 (1990).

\bibitem{Meir-PRL94} Y. Meir, Composite edge states in the $\nu $ = 2/3
fractional quantum Hall regime. Phys. Rev. Lett. 72, 2624 (1994).
\end{thebibliography}
\end{document}